\documentstyle[aps,multicol,epsf,epsfig]{revtex}

\begin{document}
\draft
\title{Entangled Webs:
Tight Bound for Symmetric Sharing of Entanglement}
\author{Masato Koashi$^1$, Vladim\'{\i}r Bu\v{z}ek$^{1,2}$,
 and Nobuyuki Imoto$^1$}
\address{$^1$CREST Research Team for Interacting Carrier Electronics,
School of
Advanced Sciences, \\
~The Graduate University for Advanced Studies (SOKEN),
Hayama, Kanagawa, 240-0193, Japan \\
$^2$Institute of Physics, Slovak Academy of Sciences,
D\'ubravsk\'a cesta 9, 842 28 Bratislava,
Slovakia}
\maketitle
\begin{abstract}
Quantum entanglement cannot be unlimitedly
shared among arbitrary number of qubits.
Larger the number of entangled pairs in an
$N$-qubit system, smaller the degree of
bi-partite entanglement is. We analyze a
system of $N$ qubits in which an arbitrary
pair of particles is entangled.  We show
that the maximum degree of entanglement
(measured in the concurrence) between any
pair of qubits is $2/N$. This tight bound
can be achieved when the qubits are prepared
in a pure symmetric (with respect to
permutations) state  with just one qubit in
the basis state $|0\rangle$ and the others
in the basis state $|1\rangle$.
\end{abstract}

\begin{multicols}{2}

Schr\"odinger
\cite{Schroedinger} has identified quantum entanglement as the key
ingredient in the paradigm of quantum mechanics. Throughout the whole
history
of modern quantum mechanics the mystery of quantum entanglement puzzled
generations of physicists \cite{Peres}.
On the other hand
in the last decade the entanglement has been
recognized as an important resource for
quantum information processing. In particular,
quantum computation \cite{Preskill},
quantum teleportation \cite{Bennett93}, quantum dense coding
\cite{Bennett92},  certain types of quantum key distributions
\cite{Ekert91} and quantum secret sharing protocols \cite{Hillery99},
are rooted in the existence of quantum entanglement.

In spite of all the progress in the understanding of the nature
of quantum entanglement there are still open questions which
have to be answered. In particular, it is not clear yet
how to quantify uniquely the degree of entanglement
\cite{BBPS,Vedral,BDSW,Hill,Horodecki00}, or how to specify
the inseparability conditions for bi-partite multi-level
systems (qudits) \cite{Kraus}. Further problem which
waits for a thorough illumination is the multiparticle entanglement
\cite{Thapliyal99}. There are several aspects of quantum
multiparticle correlations. For instance, it is the investigation
of intrinsic $n$-party entanglement (i.e, generalizations
of the GHZ state \cite{GHZ}). Another aspect of the multiparticle
entanglement is that in contrast to
classical correlation  it cannot freely be shared among many
objects.
In particular,
 Coffman {\it et al.} \cite{Coffman} have studied
a set of three qubits
$A,B$ and $C$. It has been shown that the sum
of the entanglement (measured in terms of the tangle \cite{Hill})
between the particles $AB$ and the particles $AC$ is smaller or
equal to the entanglement between the particle $A$ and the
subsystem $BC$. Wootters \cite{Wootters}
has considered an {\em infinite} collection of qubits arranged in an
open  line,
such that every pair of nearest neighbors is entangled.
In this translationally invariant
{\em entangled chain} the maximal closest-neighbor (bi-partite)
entanglement (measured in the concurrence) is bounded by the value
$1/\sqrt{2}$ (it is not known whether this bound is achievable)
\cite{Wootters}.

In this Letter, we consider a {\em finite} system of $N$ qubits in which
each pair out of $N(N-1)/2$ possible pairs is  entangled.
We show that the maximal possible bi-partite
concurrence in this case is equal to $2/N$.
The derivation of this
tight bound on the concurrence is the main result of our Letter.

The problem is formally posed as follows. Suppose that $N$ qubits,
indexed by $l=1,2,\ldots,N$, are in the state $\hat{\rho}_N$, and choose
a basis $\{|1\rangle,|0\rangle\}$ for each qubit.
We require
that the matrix form of the marginal
density operator $\rho_{ll^\prime}$
for a pair of qubits $l$ and $l^\prime$, represented in
the chosen basis, is independent  of the choice of $l$ and $l^\prime$.
Note that this requirement is satisfied if $\hat{\rho}_N$ is invariant
under any permutation of qubits. The question is to find the maximum
degree of entanglement between a pair of qubits.

It is convenient to suppose that each qubit is a spin-1/2 particle with
the spin operator $\hat{\bbox{s}}^{(l)}(l=1,\ldots N)$.
The Hilbert space of the subsystem composed of qubits 1 and 2
is a direct sum of the subspaces for the total spin 0 and 1, with
the projectors   $\hat{P}_0$ and $\hat{P}_1$ onto each subspace,
respectively. Under the condition $\hat\rho_{12}=\hat\rho_{21}
(\equiv\hat\rho)$, we have
$\hat{P}_0\hat\rho\hat{P}_1=\hat{P}_1\hat\rho\hat{P}_0=0$, since these
operators change their signs under the permutation of the two qubits.
Let us define irreducible tensors $\hat{T}^{(k)}_{j,q}$ of rank $k=0,1,2$
and components $q$ such that
$\hat{T}^{(k)}_{j,q}\equiv \sum_{m,m^\prime}\langle
k,q|1,m;1,m^\prime\rangle (\sum_{l=1}^j\hat{s}^{(l)}_{m})
(\sum_{l^\prime=1}^j\hat{s}^{(l^\prime)}_{m^\prime})$, where
$\hat{s}^{(l)}_{\pm 1}\equiv \mp(\hat{s}^{(l)}_x\pm
i\hat{s}^{(l)}_y)/\sqrt{2}$, $\hat{s}^{(l)}_{0}\equiv\hat{s}^{(l)}_z$,
and $\langle k,q|1,m;1,m^\prime\rangle$ is the Clebsh-Gordon coefficient
for forming a total spin $k$ state from two spin-1 particles.
The spin-1 part of the density operator
$\hat\rho$ can be expanded by $\hat{T}^{(k)}_{2,q}$ as
$\hat{P}_1\hat\rho\hat{P}_1=\sum_{k,q}\alpha_{k,q}\hat{T}^{(k)}_{2,q}$,
and the coefficients
$\alpha_{k,q}$ are  obtained by the relation
$\alpha_{k,q}\mbox{Tr}(\hat{T}^{(k)}_{2,-q}\hat{T}^{(k)}_{2,q})
=\mbox{Tr}(\hat{T}^{(k)}_{2,-q}\hat\rho)
=\langle \hat{T}^{(k)}_{2,-q}\rangle$, where we denote
$\mbox{Tr}(\ldots\hat\rho_N)$ as $\langle\ldots\rangle$. From the
symmetry  of $\hat\rho_N$, we have
$\langle \hat{T}^{(1)}_{N,q}\rangle
=(N/2)\langle\hat{T}^{(1)}_{2,q}\rangle$ and $\langle
\hat{T}^{(2)}_{N,q}\rangle =(N(N-1)/2)\langle\hat{T}^{(2)}_{2,q}\rangle$.
With $\hat{\rho}_N$ given, it is convenient
to choose the $x$,$y$, and $z$
axis as the principal axes for the tensor of the second-order correlation
for the total spin
$\hat{\bbox{S}}\equiv \sum_{l=1}^N\hat{\bbox{s}}^{(l)}$ of the $N$
qubits, namely,
$\langle \hat{S}_\mu\hat{S}_\nu+
\hat{S}_\nu\hat{S}_\mu\rangle=2S_\mu^2\delta_{\mu\nu}$ with
$S_\mu^2\equiv
\langle
\hat{S}_\mu^2\rangle$ (here and henceforth, suffices $\mu$ and $\nu$
represent $x,y,z$).
Matrix elements for $\hat\rho$ then takes a simple form
on the basis
$\{|{\uparrow\uparrow}\rangle-|{\downarrow\downarrow}\rangle,
|{\uparrow\uparrow}\rangle+|{\downarrow\downarrow}\rangle,
|{\uparrow\downarrow}\rangle+|{\uparrow\downarrow}\rangle,
|{\uparrow\downarrow}\rangle-|{\uparrow\downarrow}\rangle\}$ as follows,
\begin{equation}
\rho=\frac{1}{N}
\left(\begin{array}{cccc}
A_x & \langle \hat{S}_z\rangle & i\langle \hat{S}_y\rangle & 0\\
\langle \hat{S}_z\rangle & A_y & \langle \hat{S}_x\rangle & 0\\
-i\langle \hat{S}_y\rangle & \langle \hat{S}_x\rangle & A_z & 0\\
0 & 0 & 0 & A_0
\end{array}\right)
,
\label{rho}
\end{equation}
where we have introduced nonnegative parameters
$A_0\equiv\frac{N(N+2)-4\langle\hat{\bbox{S}}^2\rangle}{4(N-1)}$ and 
$A_\mu\equiv\frac{N^2-4S_\mu^2}{2(N-1)}-A_0$,
that satisfy $A_x+A_y+A_z+A_0=N$.

As a measure of entanglement between the two qubits, we use the
`concurrence' introduced by Hill and Wootters\cite{Hill}. The concurrence
$C$ can be calculated as follows.  Let $\tilde\rho$ be the time reversal
of $\rho$, which is obtained  by changing the sign of spin $\langle
\hat{\bbox{S}}\rangle$ in  the expression (\ref{rho}). The eigenvalues of
$\rho\tilde\rho$ are all real and non-negative, and let
the square roots of those be
$l_1$, $l_2$, $l_3$, and $l_4$ in decreasing
order. The concurrence $C$ is then given by
$C=\max\{l_1-l_2-l_3-l_4,0\}$.
In the present case, one of the eigenvalues of $\rho\tilde\rho$ is
$(A_0/N)^2$. Let us denote the other three as $(\lambda_1/N)^2$,
$(\lambda_2/N)^2$, and $(\lambda_3/N)^2$ in decreasing
order, and introduce parameters
$\beta\equiv\lambda_1+\lambda_2+\lambda_3$
and $\gamma\equiv\lambda_1-\lambda_2-\lambda_3$.
The concurrence is then
given by
$C=\max\{(\gamma-A_0)/N,(A_0-\beta)/N,0\}$,
where allowances are made for the possible order of $A_0$ and $\lambda_1$.

In the following, we first fix the parameters $S_\mu^2$ (hence $A_\mu$
and $A_0$), and   maximize $\gamma$ with respect to
$(X,Y,Z)\equiv
(\langle\hat{S}_x\rangle^2,\langle\hat{S}_y\rangle^2,
\langle\hat{S}_z\rangle^2)$.
We then move
$S_i^2$ to obtain the global maximum of $(\gamma-A_0)/N$, that
turns out to be, as we shall see, the maximum of the concurrence. For
simplicity, we assume that $S_z^2>S_y^2>S_x^2$ (hence
$A_z<A_y<A_x$).  The states with
some parameters equal will be considered as the limiting cases.

There are two simple bounds for the allowed values of
$(X,Y,Z)$.
One is a necessary condition for $\rho$ to be physical.
The eigenvalues for $\rho$ must be nonnegative, and the boundary
is given by the surface that satisfies $\det(\rho)=0$.
Calculating from (\ref{rho}), this surface turns out to be a plane, and
the condition for
$(X,Y,Z)$ is
\begin{equation}
f_A\equiv A_xA_yA_z-A_xX
-A_yY-A_zZ\ge0 .
\label{cond-2phys}
\end{equation}
The other one is a requirement necessary
for the spin correlations. From the inequality $\langle[\sum_i\langle
\hat{S}_i\rangle (\hat{S}_i-\langle \hat{S}_i\rangle)/S_i^2]^2\rangle\ge
0$, we obtain
\begin{equation}
f_S\equiv 1-\frac{X}{S_x^2}
-\frac{Y}{S_y^2}
-\frac{Z}{S_z^2}\ge 0.
\end{equation}
Any physical state thus falls in the region $V$, that is
defined by $f_A\ge 0$, $f_S\ge 0$, and $X,Y,Z\ge 0$

The relations that connect $(X,Y,Z)$ and $\lambda_i$ are obtained by
expanding $\det[\rho\tilde{\rho}-(\lambda^2/N^2) I]
=\prod_i(\lambda_i^2-\lambda^2)$ and equating the coefficients of
$\lambda^m$. There are three independent equations, and it is convenient
to take the following set:
\begin{eqnarray}
\label{flambda}
f_0 &\equiv& A_x^2+A_y^2+A_z^2
-2X-2Y-2Z
\nonumber
\\
&=&\lambda_1^2+\lambda_2^2+\lambda_3^2
\nonumber
\\
f_A&=&\lambda_1\lambda_2\lambda_3
\\
f_B&\equiv&-B_xB_yB_z
\left(B_x+B_y+B_z-\frac{4X}{B_x}
-\frac{4Y}{B_y}
-\frac{4Z}{B_z}
\right)
\nonumber \\
&=&\lambda_1^4+\lambda_2^4+\lambda_3^4
-2(\lambda_1^2\lambda_2^2+\lambda_2^2\lambda_3^2+\lambda_3^2\lambda_1^2)
,
\nonumber
\end{eqnarray}
where $B_\mu\equiv A_x+A_y+A_z-2A_\mu$.
The relation $A_x>A_y>A_z>0$ implies that
$B_z>B_y>|B_x|$. Note that the sign of $f_B$ is the
same as that of $\gamma$, since we can factorize as
$f_B=\beta\gamma
(\lambda_1+\lambda_3-\lambda_2)
(\lambda_1+\lambda_2-\lambda_3)$.

Let $W$ be the region defined by
$f_A\ge 0$ and $f_B\ge 0$.
As a function of $(X,Y,Z)$, $\gamma$ is continuous in
the region $W$ including the boundaries.  The gradient
$\bbox\nabla \gamma\equiv(
\partial \gamma/\partial X,\partial \gamma/\partial Y,\partial
\gamma/\partial Z)$ can formally be obtained by using the three
relations  (\ref{flambda}).
The result is
\begin{equation}
\frac{\partial \gamma}{\partial X}
=\frac{(\gamma+B_y)(\gamma+B_z)}
{\kappa},
\label{dC}
\end{equation}
where
$
\kappa\equiv2(\lambda_1-\lambda_2)
(\lambda_1-\lambda_3)
(\lambda_2+\lambda_3)\ge 0
$, and the other two are obtained by the cyclic exchange.
At the inner points of $W$, $\lambda_1>\lambda_2$ since $\gamma>0$,
and $\lambda_3>0$ since $f_A>0$. The parameter $\kappa$ is hence
positive, and the gradient $\bbox\nabla \gamma$ exists. Since $B_z>B_y>0$,
we have $\partial \gamma/\partial X>0$ for the inner points of $W$.

The geometry of $W$ and $V$ is derived as follows.
Let the points on which the planes $f_B=0$, $f_A=0$, and $f_S=0$ intersect
$X$-axis be $P_{BX}$, $P_{AX}$, and $P_{SX}$, respectively, and denote
other six points on $Y$- and $Z$-axis similarly. The relation
$A_xA_y-(B_x+B_y+B_z)B_z/4=B_xB_y/4$, and the ones obtained
by the cyclic exchange of  $\{x,y,z\}$, tells us the following.
When $B_x>0$, the triangle
$\pi_B\equiv P_{BX}P_{BY}P_{BZ}$ does not intersect with
$\pi_A\equiv P_{AX}P_{AY}P_{AZ}$, and
lies closer to
the  origin. $W$ is the sandwiched region of the two triangles.
The triangle
$\pi_S\equiv P_{SX}P_{SY}P_{SZ}$ may intersect
with $\pi_A$ and $\pi_B$ or not. When $B_x<0$,
$f_B>0$ (hence $\gamma>0$) is satisfied by all the points that satisfy
$f_A\ge 0$. When $B_x=0$, $\gamma=0$ for the points on $YZ$-plane,
and $\gamma>0$ for $X>0$. Combining these observation with
$\partial \gamma/\partial X>0$, we conclude
that in $V$, $\gamma$ takes its maximum on points on the
boundaries $\pi_A$ or $\pi_S$, and never on the inner points.

To determine the behavior of $\gamma$ on the boundaries, we first derive
the value of $\gamma$ on the axes explicitly.
For the points on $Z$-axis and satisfy
(\ref{cond-2phys}),
the roots $\{\lambda_i\}$ are found to be
$\left\{A_z,\left[
\sqrt{(A_x+A_y)^2-4\langle \hat{S}_z\rangle^2}
\pm(A_x-A_y)\right]/2\right\}.
$
The expression for $\gamma$ depends on which is the largest root.
The roots for the points on the other two axes are similarly obtained.
Applying these for the vertices of $\pi_A$, we have
$\gamma(P_{AX})=B_y$ and $\gamma(P_{AY})=\gamma(P_{AZ})=|B_x|$.
On the vertices of $\pi_S$, $\gamma$ is well-defined
only when they are in the region $W$. When $P_{SZ}$ is in $W$,
\begin{eqnarray}
0&\le& 4(A_xA_y-S_z^2)
\nonumber
\\
&=&(A_z+A_0)(A_z+A_0-2)-(A_x-A_y)^2,
\end{eqnarray}
and hence $A_z>2-A_0$.
Then $\gamma$ is written as
\begin{eqnarray}
\gamma(P_{SZ})&=&
A_z-\sqrt{(A_x+A_y)^2-4S_z^2}
\nonumber
\\
&=&
A_z-\sqrt{(A_z+A_0)(A_z+A_0-2)}
,
\label{psz}
\end{eqnarray}
This is a decreasing function of $A_z$ and noting $S_z^2>0$, we have
$2-A_0>\gamma(P_{SZ})>N/(N-1)-A_0\equiv\gamma_m$. Similarly,
$\gamma(P_{SX})>\gamma_m$ and $\gamma(P_{SY})>\gamma_m$ if those points
are in $W$.

 Next, we will evaluate
the gradient on the boundaries.
On the boundary $\pi_A$, $\kappa$ is not always positive, but
$\kappa=0$ is possible only for the following two cases.
(a) $\lambda_1=\lambda_2$.  This occurs only when $f_A=f_B=0$,
or equivalently, on the segment $P_{AY}P_{AZ}$ when $B_x=0$.
(b) $\lambda_2=\lambda_3=0$. This occurs if and only if $f_A=0$ and
$(f_0)^2-f_B=0$. The intersection of $f_A=0$ and
$(f_0)^2-f_B=0$ is a parabola. The point
$P_0(A_z^2(A_x-A_y)/(A_x-A_z),0,A_x^2(A_y-A_z)/(A_x-A_z))$ is on this
parabola. At $P_0$, $\bbox{u}=(A_x,A_y,A_z)$ is normal to $f_A=0$, and
$\bbox{v}=(-A_xB_y,A_yB_y-2A_xA_z,-A_zB_y)$ is normal to $(f_0)^2-f_B=0$.
Since $\bbox{u}\times\bbox{v}$ has a vanishing $Y$ component,
the parabola is tangent to $XZ$-plane at $P_0$.
Another point $(A_y^2(A_x-A_z)/(A_x-A_y),-A_x^2(A_y-A_z)/(A_x-A_y),0)$,
on which $Y<0$, is also on the parabola. We thus conclude that
the gradient exists on $\pi_A$ except for the segment $P_{AY}P_{AZ}$
and $P_0$.

Let us define three particular directions on $\pi_A$ as
$\bbox{q}^{yx}\equiv(A_yA_z,-A_zA_x,0)$,
$\bbox{q}^{zx}\equiv(A_yA_z,0,-A_xA_y)$, and
$\bbox{q}^{yz}\equiv(0,-A_zA_x,A_xA_y)$.
The differential coefficients of $\gamma$ for these directions are
calculated to be
$\bbox{q}^{yx}\cdot\bbox{\nabla}\gamma=
A_z(B_z^2-\gamma^2)
(A_x-A_y)/\kappa$,
$\bbox{q}^{zx}\cdot\bbox{\nabla}\gamma=
A_y(B_y^2-\gamma^2)
(A_x-A_z)/\kappa$, and
$\bbox{q}^{yz}\cdot\bbox{\nabla}\gamma=
A_x(\gamma^2-B_x^2)
(A_y-A_z)/\kappa$.
Noting that
$\gamma(P_{AX})=B_y$ and $\gamma(P_{AY})=\gamma(P_{AZ})=|B_x|$,
we conclude that $\bbox{q}^{yx}\cdot\bbox{\nabla}\gamma>0$,
$\bbox{q}^{zx}\cdot\bbox{\nabla}\gamma>0$ and
$\bbox{q}^{yz}\cdot\bbox{\nabla}\gamma>0$ on $\pi_A$ except
for on the segment $P_{AY}P_{AZ}$ and $P_0P_{AX}$.
On $P_0P_{AX}$, $\bbox{q}^{zx}\cdot\bbox{\nabla}\gamma=0$,
so that $\gamma$ is constant.

Similarly, define directions on $\pi_S$ as
$\bbox{p}^{xy}\equiv(-S_x^2,S_y^2,0)$
$\bbox{p}^{yz}\equiv(0,-S_y^2,S_z^2)$
and
$\bbox{p}^{xz}\equiv(-S_x^2,0,S_z^2)$.
Let $\pi_{SV}$ be the intersection of $\pi_S$ and $W$.
On $\pi_{SV}$, we have
$
\bbox{p}^{xy}\cdot\bbox{\nabla}\gamma=
(\gamma+B_z)
(\gamma-\gamma_m)
(S_y^2-S_x^2)/\kappa
$
,
$
\bbox{p}^{xz}\cdot\bbox{\nabla}\gamma=
(\gamma+B_y)
(\gamma-\gamma_m)
(S_z^2-S_x^2)/\kappa
$, and
$
\bbox{p}^{yz}\cdot\bbox{\nabla}\gamma=
(\gamma+B_x)
(\gamma-\gamma_m)
(S_z^2-S_y^2)/\kappa
$.
When $B_x<0$, $\gamma(P_{AY})=\gamma(P_{AZ})=|B_x|$ and (\ref{dC})
implies that
$\gamma=|B_x|$ on $YZ$-plane, and $\gamma+B_x>0$ for $X>0$.
Since we have seen that $\gamma>\gamma_m$ on the vertices on $\pi_{SV}$,
$\bbox{p}^{xz}\cdot\bbox{\nabla}\gamma>0$ and
$\bbox{p}^{xy}\cdot\bbox{\nabla}\gamma>0$ everywhere on $\pi_{SV}$,
and $\bbox{p}^{yz}\cdot\bbox{\nabla}\gamma>0$ on $\pi_{SV}$ except for
segment $P_{SY}P_{SZ}$.

\begin{figure}
\centerline {\epsfig{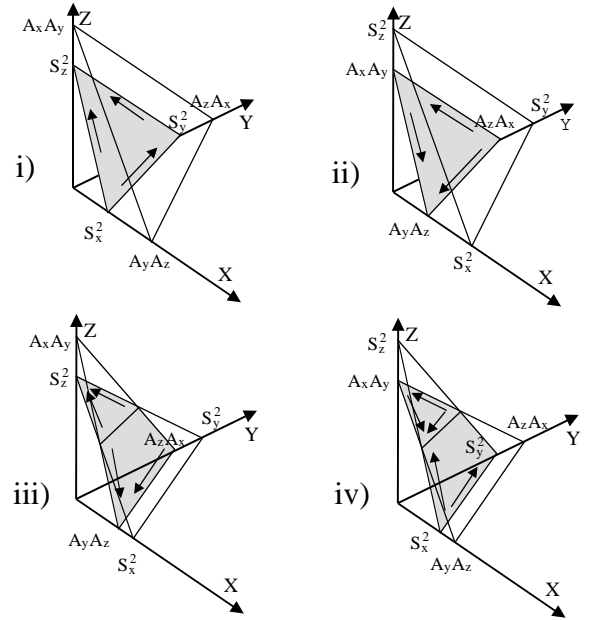}}
\caption{Allowed regions for the parameters $\{X,Y,Z\}$.
The shaded surfaces are boundaries for the physically allowed
states. The arrows show the directions in which the concurrence
increases.
\label{f1}
}
\end{figure}

Now we are in a position to find the maximum of $\gamma$. We must consider
the following four cases separately (see Fig.~\ref{f1}).

i) $A_xA_y\ge S_z^2$ and $A_yA_z>S_x^2$.
In this case, $A_z>2-A_0$ is necessary.
$\gamma$ takes its maximum on $P_{SZ}$ and the value is given by
(\ref{psz}) and $\gamma<2-A_0$.
$\gamma$ approaches $2$ only in the limit of
$A_0\rightarrow 0$, $A_z\rightarrow 2$,
$B_x\rightarrow 2$,
and $B_y\rightarrow 2$.
 This limit can be taken only if
$N\ge 6$, since $A_y\ge A_z$ must hold in the limit.
When $N>6$,
$P_{SZ}$ is the only point that attains $\gamma=2$
since $\bbox{p}^{yz}\cdot\bbox{\nabla}\gamma>0$ and
$\bbox{p}^{xz}\cdot\bbox{\nabla}\gamma>0$ still hold in
the limit.
When $N=6$, $\gamma=2$ everywhere on $\pi_{A}$, but these states
are equivalent in the sense that they are related by the rotation
of the whole system.

ii) $A_xA_y< S_z^2$ and $A_yA_z\le S_x^2$. In this case,
from the relation
$0\le 4(S_x^2-A_yA_z)=-(B_y+A_0-2)(B_y+A_0)-(B_z-B_y)(B_y+A_0-1)$,
$B_y<2-A_0$ is necessary.
$\gamma$ takes its maximum on $P_0P_{AX}$ and the value is
$\gamma=B_y<2-A_0$.
$\gamma$ approaches $2$ only in the limit of
$A_0\rightarrow 0$, $A_x\rightarrow 2$,
$B_y\rightarrow 2$,
and $B_z\rightarrow 2$. This limit can be taken only if
$N\le 6$, since $A_x\ge A_y$ must hold in the limit.
When $N<6$, $P_0$ coincides with $P_{AX}$ in the limit
and $P_{AX}$ is thus the only point that attains $\gamma=2$.
When $N=6$, the limit is the same as in the case i).

iii) $A_xA_y\ge S_z^2$ and $A_yA_z\le S_x^2$. The maximum
is the larger of $\gamma(P_{AX})$ and $\gamma(P_{SZ})$.
Depending on $N$, one of them or both can approach
$2$. The limit is the same as described in the cases i) and ii).

iv) $A_xA_y<S_z^2$ and $A_yA_z>S_x^2$.
In this case, from the relation
$C(P_{AX})>C(P_{SX})$, we have $B_y>\gamma_m$.
We also have $A_0<2$ since $(2-A_0)(B_y+A_0+2A_z)=
4(S_z^2-A_xA_y)+2A_z+B_y(N-2+4S_x^2)/(N-1)>0$.
$\gamma$ takes its maximum at $P_1$, that is the intersection of
$P_{AZ}P_{AX}$  and $P_{SZ}P_{SX}$. Since $\lambda_3=0$ at $P_1$,
$\gamma=\lambda_1-\lambda_2$ and $\beta=\lambda_1+\lambda_2$.
Then we have
$\beta^2+\gamma^2=2f_0$
and
$\beta^2\gamma^2=f_B$.
This implies that $\gamma^2$ is the smaller of the two roots
$t=t_\alpha, t_\beta$
of the equation $t^2-2f_0t+f_B=0$. The coordinates of $P_1$
can explicitly be obtained by solving $f_A=0$ and $f_S=0$ with $Y=0$.
Substituting the result into $f_0$ and $f_B$ in the equation of $t$,
we finally obtain $t_\alpha=B_y^2$ and
\end{multicols}
\vspace{-0.5cm}
\widetext
\begin{eqnarray}
t_\beta&=&
\frac{4A_xA_z}{(B_y-\gamma_m)(N-1)}+A_0(A_0-2)
\nonumber \\
&=&(2-A_0)^2
+2A_0\frac{B_y+A_0-2}{B_y-\gamma_m}
-\frac{2(B_y+A_0-2)(B_y+N)+(B_z-B_y)(B_y-B_x)}{(N-1)(B_y-\gamma_m)}.
\end{eqnarray}
\begin{multicols}{2}
When $B_y\ge 2-A_0$, we have $t_\beta<4-A_0(2-A_0)<4$.
We thus conclude that $\gamma<2$ for all values of $B_y$.
When $N<6$,
$\gamma$ approaches $2$ only in the limit of $A_0\rightarrow 0$,
$B_y\rightarrow 2$
, $B_z\rightarrow 2$,
and $A_x\rightarrow 2$.
$P_1$ approaches $P_{AX}$ in this limit, so the limit
is the same as in the case ii).
When $N>6$,
$\gamma\rightarrow 2$ only in the limit of $A_0\rightarrow
0$, $B_y\rightarrow 2$,
$B_x\rightarrow 2$,
$A_z\rightarrow 2$, and $P_1\rightarrow P_{SZ}$. This is
the same limit as in the case i). When $N=6$,
$\gamma\rightarrow 2$ in the limit of $A_0\rightarrow
0$ and all $A_\mu$ approaching $2$.

Finally, we show that $A_0-\beta$ is smaller than 2. In $W$, $f_0$
takes its minimum on $P_{AZ}$, and its value is $(A_x-A_y)^2+A_z^2$.
We thus have $A_0-\beta\le A_0-\sqrt{f_0}\le A_0-A_z
=(N-2S_x^2-2S_y^2)/(N-1)<N/(N-1)<2$ for $N\ge 3$.

Combining  all cases, we reach a conclusion that the
maximum value of the concurrence is $2/N$, and this value is
reached only by the state satisfying the constraints
$\langle \hat{S}_z^2 \rangle=(N/2-1)^2$,
$\langle \hat{S}_z \rangle=N/2-1$,
$\langle \hat{S}_x^2 \rangle=\langle \hat{S}_y^2 \rangle
=(3N-2)/4$,
and $\langle \hat{S}_x \rangle=\langle \hat{S}_y \rangle=0$,
if $z$-axis is suitably chosen. Such a state exists -
it is the eigenstate
of $\hat{S}_z$ with eigenvalue $N/2-1$ with total spin $N/2$.
This state is equally-weighted  in-phase superposition of any one qubit
being in the state $|0\rangle$ and the other $N-1$ qubits in the
state $|1\rangle$. This is a
permutationally invariant pure
 state that  highly (in the order of $N$) breaks the
symmetry between the basis states $|0\rangle$ and $|1\rangle$.

It is worth noting that the optimal state for our problem was
found to be a pure state. This
is not trivial because the symmetry is required for  density
operators. Obviously,  possible mixed states span a larger Hilbert
space than that spanned by possible pure states.

For entangled chains,
the best state so far is a {\em mixed state}
\cite{Wootters}. It will be
interesting to investigate  entangled chains composed of
 a {\em finite} number of qubits.
Specifically, instead of an infinite line of qubits one can consider
an {\em entangled} loop composed of even number of qubits $2N$
 (here we assume $N>1$
because for $N=1$ we would have a two-qubit ``loop'' which is maximally
entangled when prepared in a Bell state, in this case the concurrence
is equal to unity).
One can find a pure $2N$-qubit
rotationally-invariant state for which
the closest-neighbor  bi-partite entanglement has the concurrence
 $C=(2+2^{N-2})/(2+2^{N})$. In the limit
$N\rightarrow\infty$ this concurrence is $C_\infty = 1/4$. We conjecture
that this is the maximal value of the concurrence for {\em pure} infinite
entangled chains. This means that impure states in the limit of
infinite number of particles may attain higher bi-partite entanglement
than pure states.

This work was supported by a Grant-in-Aid for Encouragement of Young
Scientists (Grant No.~12740243) by Japan Society of the Promotion of
Science, and
by the IST project EQUIP under the contract IST-1999-11053.

\end{multicols}

\begin{thebibliography}{99}

\bibitem{Schroedinger}
    E.~Schr\"{o}dinger,
       Naturwissenschaften {\bf 23}, 807,   (1935);
 {\it ibid.} {\bf 23}, 823 (1935);
 {\it ibid.} {\bf 23}, 844 (1935).


\bibitem{Peres}
     A.~Einstein, B.~Podolsky, and N.~Rosen,
        Phys. Rev. A {\bf 47}, 777 (1935);
     J.S.Bell,
        Physics {\bf 1}, 195 (1964);
     A. Peres,
        {\it Quantum Theory: Concepts and Methods}
        (Kluwer, Dordrecht, 1993).

\bibitem{Preskill}
     J.~Gruska,
        {\it Quantum Computing}
        (McGraw-Hill,1999);
     J.~Preskill,
         {\it Quantum   Theory   Information and} {\it  Computation}
         ({\tt www.theory.caltech.edu/people/preskill}).

\bibitem{Bennett93}
     C.~H.~Bennett, {\it et al.},
        {Phys. Rev. Lett.} {\bf 70}, 1895 (1993).


\bibitem{Bennett92}
    C.~H.~Bennett and S.~Wiesner,
         {Phys. Rev. Lett.} {\bf 69}, 2881 (1992).

\bibitem{Ekert91}
    A.~K.~Ekert,
         {Phys. Rev. Lett.} {\bf 67}, 661 (1991).

\bibitem{Hillery99}
    M.~Hillery, {\it et al.},
          {Phys. Rev. A} {\bf 59}, 1829 (1999).


\bibitem{BBPS}
  C.~H.~Bennett, {\it et al.},
       {Phys.~Rev.~A} {\bf 53}, 2046 (1996).

\bibitem{Vedral}
  V.~Vedral, {\it et al.},
       {Phys. Rev. Lett.} {\bf 78}, 2275 (1997);
  V.~Vedral, {\it et al.},
       {Phys. Rev. A} {\bf 56}, 4452 (1997);
  V.~Vedral and M.~B.~Plenio,
       {Phys. Rev. A} {\bf 57}, 1619 (1998).

\bibitem{BDSW}
  C.~H.~Bennett, {\it et al.},
       {Phys.~Rev.~A} {\bf 54}, 3824 (1996).

\bibitem{Hill}
  S.~Hill and W.~K.~Wootters,
       {Phys. Rev. Lett.} {\bf 78}, 5022 (1997);
  W.~K.~Wootters,
        Phys. Rev. Lett. {\bf 80}, 2245 (1998).

\bibitem{Horodecki00}
    M.~Horodecki, {\it et al.},
          {Phys. Rev. Lett.} {\bf 84}, 2014 (2000).

\bibitem{Kraus}
    B.~Kraus, {\it et al.},
    {\tt arXiv quant-ph/9912010} (1999);
   P.~Rungta {\it et al.}
    {\tt arXiv quant-ph/0001075} (2000).

\bibitem{Thapliyal99}
   A.~V.~Thapliyal,
      Phys. Rev. A {\bf 59}, 3336 (1999);
   J.~Kempe,
      Phys. Rev. A {\bf 60}, 910 (1999).

\bibitem{GHZ}
   D.~M.~Greenberger, {\it et al.},
   Am. J. Phys. {\bf 58}, 1131 (1990).



\bibitem{Coffman}
  V. Coffman, {\it et al.},
        {\tt arXiv quant-ph/9907047} (1999).


\bibitem{Wootters}
  W.~K. Wootters,
        {\tt arXiv quant-ph/0001114} (2000).

\end{thebibliography}
\end{document}